\documentstyle[12pt]{article}
\def\be{\begin{equation}}
\def\ee{\end{equation}}
\def\bea{\begin{eqnarray}}
\def\eea{\end{eqnarray}}
\def\lsim{\:\raisebox{-0.5ex}{$\stackrel{\textstyle<}{\sim}$}\:}
\def\gsim{\:\raisebox{-0.5ex}{$\stackrel{\textstyle>}{\sim}$}\:}
\begin{document}

\textwidth=16cm
\textheight=25.5cm
\topmargin=-3.5cm
\abovedisplayskip=3pt
\belowdisplayskip=3pt

\begin{flushright}
TIFR/TH/93-48
\end{flushright}
\bigskip
\begin{center}
\large Like Sign Dilepton Signature for Gluino Production at LHC with or
without R Conservation \\
\bigskip
\bigskip
H. Dreiner$^{a^\dagger}$, Manoranjan Guchait$^b$, D.P.
Roy$^{c^{\dagger\dagger}}$ \\
\end{center}
\bigskip
\bigskip

$^a$Physics Department, University of Oxford, Oxford OX1 3RH, UK \\

$^b$Physics Department, Jadavpur University, Calcutta - 700 032, India \\

$^c$Theory Group, Tata Institute of Fundamental Research, Bombay - 400
005, India \\

\bigskip
\bigskip

\begin{center}
\underbar{Abstract} \\
\end{center}
\smallskip

The isolated like sign dilepton signature for gluino production is
investigated at the LHC energy for the $R$ conserving as well as the $L$ and
$B$ violating SUSY models over a wide range of the parameter space.  One
gets viable signals for gluino masses of 300 and 600 GeV for both $R$
conserving and $L$ violating models, while it is less promising for the
$B$ violating case.  For a 1000 GeV gluino, the $L$ violating signal
should still be viable; but the $R$ conserving signal becomes too small at
least for the low luminosity option of LHC.

\bigskip
\bigskip

\hrule width 5cm
\smallskip

\noindent $^\dagger$Present Address: Institute of Theoretical Physics,
ETH, Zurich, Switzerland.

\noindent $^{\dagger\dagger}$E-Mail: DPROY@TIFRVAX.BITNET

\newpage

\noindent I. \underbar{\bf Introduction} \\
\bigskip

The hadron colliders offer by far the best discovery limit for
superparticles because of their higher energy reach.  The superparticles
having the largest production cross section at hadron colliders are the
strongly interacting ones -- the squark $\tilde q$ and gluino $\tilde g$.
Therefore there has been a good deal of discussion on the search of these
superparticles at the present and proposed hadron colliders [1,2].  So far
the search programme has been largely based on the missing $p_T$ signature
assuming $R$ conservation [3].  The latter implies pair production of
superparticles followed by their decay into the lightest superparticle
(LSP) which has to be stable.  It is also required to be colourless and
neutral for cosmological reasons [4].  The LSP escapes detection due to
its feeble interaction with matter resulting in the missing $p_T$
signature for superparticle production.

There is a growing realisation in recent years however that the
multilepton signature, and in particular the like sign dilepton (LSD)
signature, may play an equally important role in superparticle search for
the following reasons.  1) The squark and gluino searches are expected to
be carried over the mass range of several hundred GeV at LHC/SSC.  The
dominant decay mode for a squark or gluino in this mass range is not its
direct decay into the LSP, which is generally assumed to be the lightest
neutralino, but a cascade decay via the heavier neutralino and chargino
states.  The cascade decay proceeds through the emission of $W$ or $Z$
which have significant leptonic branching ratios.  Thus one expects two
(or more) leptons resulting from the cascade decay of the squark or gluino
pair.  More over, in the latter case the two leptons are expected have
like sign half the time due to the Majorana nature of gluino [5-7].  2) In
the $R$ violating $(R\!\!\!/)$ SUSY models there is no missing $p_T$
signature, since the LSP is no longer stable.  Instead it decays into a
leptonic or baryonic channel depending on whether the $L$ or $B$ violating
Yukawa coupling is the dominant one [8-10].  In the former case one
expects two (or more) leptons resulting from the decay of the LSP pair.
Again the two leptons are expected to have like sign half the time due to
the Majorana nature of the LSP.  In the latter case there is no viable
signature from the baryonic decay of LSP; the like sign dileptons coming
from the cascade decay process provides by far the best signature for this
case [9].

Thus there are two contributions to the like sign dilepton signature for
gluino production -- 1) from the cascade decay of gluino into LSP, which
holds for $R$ conserving as well as the $L\!\!\!/$ and $B\!\!\!/$ SUSY
models, and 2) from the leptonic decay of LSP in the $L\!\!\!/$ moel.  In
the latter case the size of the like sign dilepton signal is expected to
be large.  In fact the Tevatron dilepton data is already known to give a
gluino (squark) mass limit in the $L\!\!\!/$ SUSY model [10], which is as
large as that obtained from the corresponing missing-$p_T$ data for the
$R$ conserving case [11].  On the other hand the LSD signal arising from
the cascade decay process is suppressed by the leptonic branching
fractions of the two vector bosons, and hence expected to be relatively
small in size.  Nonetheless it is expected to provide a useful
alternative signature for gluino production in the $R$ conserving SUSY
model, since it has a smaller background compared to the missing-$p_T$
channel.  Moreover it provides the only signature for gluino production in
the $B\!\!\!/$ SUSY model as mentioned above.  Thus it is important to
make a systematic study of the LSD signal, arising from both these
sources, for the gluino mass range of interest to LHC/SSC along with the
corresponding background.  The present work is devoted to this excercise.

To be specific we shall concentrate on the LHC energy of
\be
\sqrt {s} = 16 ~{\rm TeV},
\ee
and assume a typical luminosity of 10 events/fb corresponding to the low
luminosity option of LHC.  Of course any viable signal here shall be even
more viable at SSC or the high luminosity option of LHC.  We shall study
gluino production and decay under the assumption
\be
m_{\tilde g} < m_{\tilde q}
\ee
in which case they provide the most important signal for superparticle
search at hadron colliders.  This inequality seems to be favoured by a
large class of SUSY models.  However, we shall briefly discuss how the
results would change if the squarks are lighter than the gluino, in which
case the dominant superparticle signal would come from the production and
decay of squarks.

The paper is organised as follows.  The standard model (SM) background for
the like sign dilepton channel is briefly discussed in section II.
Section III gives the formalism of gluino cascade decay into the LSP via
the heavier neutralino and chargino states.  It tabulates the masses and
the compositions of the neutralino and chargino states, resulting from the
diagonalisation of their mass matrices, for a wide range of gluino mass
and the other SUSY parameters.  It also gives the branching fractions of
gluino decay into these states.  Section IV describes the LSP decay in the
$R\!\!\!/$ SUSY models.  Section V compares the resulting like sign
dilepton signals with the SM background.  The main conclusions are
summarised in section VI.
\bigskip

\noindent II. \underbar{\bf Standard Model Background for the Like Sign
Dilepton Channel} \\
\smallskip

\nobreak
The SM background to the like sign dilepton channel at LHC has been
studied in detail in [12].  We shall only summarise the essential points
here.  The two main sources of LSD background are from
\be
gg \rightarrow b\bar b ~(b\bar bg)
\ee
via $B\bar B$ mixing, and
\be
gg \rightarrow t\bar t ~(t\bar tg)
\ee
follwed by the sequencial decay of one of the $t$ quarks into $b$.  In the
first case both the leptons originate from $B$ particle decay, $B
\rightarrow \ell \nu D ~(D^\star)$, while in the second case one of the
leptons (the softer one) originates from $B$.  Consequently both the
contributions can be suppressed by imposing isolation cut on the leptons
[13].  Moreover the isolation cut for the $B$ decay lepton is known to
become more powerful with the lepton $p_T$, resulting in a $p_T$ cutoff
for the isolated lepton [14]
\be
p^\ell_T \lsim E^{AC}_T \left({m^2_B - m^2_{D(D^\star)} \over
m^2_{D(D^\star)}}\right).
\ee
Substituting for the bottom and charm particle masses one ses that a
typical isolation cut of
\be
E^{AC}_T < 10 ~{\rm GeV}
\ee
at LHC [15] implies a lepton $p_T$ cutoff
\be
p^\ell_T \lsim 60~{\rm GeV}.
\ee
Loss of visible $E^{AC}_T$ due to semileptonic $D$ decay and energy
resolution leads to a small spill over of the isolated lepton background
beyond this kinematic cutoff.  On the other hand the contributions to
$E^{AC}_T$ from the fragmentation of $b$ quark into $B$ particle as well
as the underlying
event tend to strengthen the bound.  All these effects are taken into
account in the ISAJET [16] Monte Carlo calculation of this background in
[12], which shows that the background becomes negligibly small beyond the
lepton $p_T$ of 60 GeV.  This is evidently a powerful result, which can be
exploited in the search of the gluino signal in the LSD channel.  We shall
use this result in our analysis.  We shall also include the fake LSD
background arising from the misidentification of one of the lepton charges
in $t\bar t \rightarrow \ell^+\ell^- X$.
\bigskip

\noindent III. \underbar{\bf Cascade Decay of Gluino into LSP} \\
\bigskip

\nobreak
We shall work within the framework of the minimal supersymmetric standard
model (MSSM) so as to have the minimum number of parameters [1,2].  The
gluino cascade decay into the LSP proceeds via the heavier neutralino and
chargino states.  There are 4 neutralino states, which are mixtures of the
4 basic interaction states, i.e.
\be
\chi^0_i = N_{i1} \tilde B + N_{i2} \tilde W^3 + N_{i3} \tilde H^0_1 +
N_{i4} \tilde H^0_2.
\ee
The masses and compositions of the neutralinos are obtained by
diagonalising the mass matrix [1,2,5,17]
\be
M_N = \left(\matrix{M_1 & 0 & -m_Z \sin\theta_W\cos\beta & m_Z
\sin\theta_W \sin\beta \cr 0 & M_2 & m_Z\cos\theta_W\cos\beta &
-m_Z\cos\theta_W \sin\beta \cr -m_Z\sin\theta_W\cos\beta & m_Z\cos\theta_W
\cos\beta & 0 & -\mu \cr m_Z\sin\theta_W \sin\beta & -m_Z\cos\theta_W
\sin\beta & -\mu & 0}\right).
\ee
where $M_1$ and $M_2$ are the soft masses of the bino $\tilde B$ and wino
$\tilde W$ respectively, $\mu$ is the supersymmetric higsino mass
parameter and $\tan\beta$ is the ratio of the two higgs vacuum
expectation values.  The two soft gaugino masses are related to that of
the gluino in the MSSM, i.e.
\be
M_2 = {\alpha \over \sin^2\theta_W\alpha_s} \cdot m_{\tilde g} \simeq 0.3
{}~m_{\tilde g}
\ee
\be
M_1 = {5 \over 3} \tan^2\theta_W ~~~M_2 \simeq 0.5 ~M_2.
\ee
Thus there are 3 independent parameters, $m_{\tilde g}$, $\mu$ and
$\tan\beta$, defining the mass matrix.  The Majorana nature of the
neutralinos ensures that the mass matrix is in general complex
symmetric and hence can be
diagonalised by only one unitary matrix $N$, i.e.
\be
N^\star M_N N^{-1} = M^D_N.
\ee
We have followed the analytical prescription of diagonalising this mass
matrix recently suggested in [18].  But we have also cross-checked our
results extensively with the numerical diagonalisation program EISCH1.FOR
of CERN library as well as the published results of [2,5,19].

The two chargino mass states are mixtures of the charged wino $\tilde
W^\pm$ and Higgsino $\tilde H^\pm$.  Their masses and compositions are
obtained by diagonalising the corresponding chargino mass matrix
\be
M_C = \left(\matrix{M_2 & \sqrt{2} m_W \sin\beta \cr \sqrt{2} m_W
\cos\beta & \mu}\right).
\ee
This is done via the biunitary transformation
\be
U ~M_C~V^{-1} = M^D_C
\ee
where $U$ and $V$ are $2 \times 2$ unitary matrices, which diagonalise the
hermitian (real symmetric) matrices $M_C M^\dagger_C$ and $M^\dagger_C
M_C$ respectively.  Explicit expressions for $U$ and $V$ may be found in
[1,5] along with those of the mass eigenvalues.  The corresponding chargino
eigenstates are
\bea
\chi^\pm_{iL} &=& V_{i1} \tilde W^\pm_L + V_{i2} \tilde H^\pm_L \nonumber
\\[2mm] \chi^\pm_{iR} &=& U_{i1} \tilde W^\pm_R + U_{i2} \tilde H^\pm_{R}
\eea
where $L$ and $R$ refer to the left and right handed helicity states.  We
shall use the real orthogonal representation for the unitary matrices
$U,V$ and $N$.  Moreover the chargino and neutralino eigenstates shall be
labelled in increasing order of mass, with
\be
\chi^0_1 \equiv \chi
\ee
representing the LSP.

Table I (a,b,c) show the masses and compositions of the neutralino and
chargino states for three representative values of the gluino mass which
are of interest to LHC/SSC; i.e.
\be
m_{\tilde g} = 300,600 ~{\rm and}~ 1000~{\rm GeV}.
\ee
The results are not very sensitive to the variation of $\tan\beta$ over
the range allowed by MSSM, i.e. $1 < \tan\beta < m_t/m_b ~(\simeq 30)$.
We have chosen 2 representative values
\be
\tan\beta = 2 ~{\rm and}~ 10;
\ee
the current lower mass bounds of top quark and neutral higgs boson do not
seem to favour $\tan\beta = 1$ [2].  On the other hand, the results are
quite sensitive to the variation of $\mu$ over the range $-M_2 < \mu <
M_2$.  Therefore we have chosen 5 representative values of this variable,
i.e.
\be
\mu = 0.1 m_W, ~\pm m_W, ~\pm 4 m_W
\ee
as in ref. [5].  This also helps us cross-check some of our results with
theirs [20].  The SM parameters used are
\be
m_W = 80 ~{\rm GeV}, ~m_Z = 91 ~{\rm GeV}, ~\sin^2\theta_W = 0.233,
{}~\alpha = 1/128, ~\alpha_s = 0.115.
\ee

The masses and compositions of the neutralinos are shown in the 2nd and
3rd columns of Table I, while those of the charginos are shown in the 5th
and 6th columns.  The upper and lower entries of 6th column refer to the
compositions of the left and right handed components of the chargino
respectively.  The sign of a mass affects the phases of the corresponding
couplings [5]; these are irrelevant however in the approximation we shall
be working in.  One should note the following systematics in the masses and
compositions of the neutralino and chargino states.

\begin{enumerate}

\item[{1)}] For $\mu = \pm 4~m_W$, the higgsino mass parameter is
generally larger than $M_1$ and $M_2$.  Consequently the LSP
$(\chi^0_1)$ is dominated by the bino $\tilde B$ component, while the
second lightest neutralino $\chi^0_2$ and the lightest chargino
$\chi^\pm_1$ are dominated by the wino $\tilde W$.  These are clearly
reflected in their masses and compositions.  Only at $m_{\tilde g} = 1000$
GeV, the $\chi^0_2$ and $\chi^\pm_1$ acquire significant higgsino
components as $M_2 \simeq |\mu|$.  Evidently the gaugino dominance of
$\chi^0_1$, $\chi^0_2$ and $\chi^\pm_1$ is expected to hold even better at
$|\mu| > 4 ~M_W$.

\item[{2)}] For $\mu = -m_W$, the higgsino mass parameter is comparable to
$M_1,M_2$ at $m_{\tilde g} = 300$ GeV and smaller at $m_{\tilde g} = 600$
and 1000 GeV.  Consequently the $\chi^0_1, \chi^0_2$ and $\chi^\pm_1$
contain significant admixtures of the gaugino and higgsino components at
$m_{\tilde g} = 300$ GeV, while they are dominated by the higgsinos at
$m_{\tilde g} = 600$ and 1000 GeV.

\item[{3)}] For $\mu = m_W$ and $0.1~m_W$, the $\chi^0_1, ~\chi^0_2$
and $\chi^\pm_1$ have large higgsino components as expected.  However,
this part of the parameter space is disallowed by the LEP data, which
gives a lower mass limit of $\sim m_Z/2$ for the lightest chargino as well as
higgs dominated neutralino [3,21].  Only at $m_{\tilde g} = 1000$ GeV does
the $\mu = m_W$ value falls marginally within the allowed region.
\end{enumerate}

Thus we see that the values of $\mu = \pm 4 ~m_W$ and $-m_W$ are generally
representative of the two extreme cases where the lighter neutralino and
chargino states are dominated by the gaugino and higgsino components
respectively, while it is the opposite for the heavier ones.  Note that
for $m_{\tilde g} = 300$ GeV the lighter states have substantial gaugino
components even at $\mu = -m_W$.  Nonetheless the latter represents the
extreme composition as it lies close to the boundary of the experimentally
allowed region [3,19].

We shall estimate the branching fractions of gluino decay into the above
chargino and neutralino states by neglecting the contributions of top
quark $(\tilde g \rightarrow t\bar t \chi^0_i, t\bar b \chi^-_i)$ as well
as the loop induced processes $(\tilde g \rightarrow g \chi^0_i)$.  This
seems to be a reasonable approximation for the bulk of the parameter space
under investigation [2].  Thus the decay processes of interest are
\be
\tilde g {\buildrel {\tilde q} \over \rightarrow} q \bar q \chi^0_i
\ee
\be
\tilde g {\buildrel {\tilde q} \over \rightarrow} q' \bar q \chi^\pm_i
\ee
where $q$ and $q'$ are understood to represent the light quarks of a given
generation.  The relevant interaction terms for these processes are
\bea
{\cal L}_{q\tilde q\tilde g} &=& {ig_s \over \sqrt{2}} \tilde q_L^\dagger
\bar{\tilde g}_A \lambda_A q_L + {ig_s \over \sqrt{2}} \tilde q^\dagger_R
\bar{\tilde g}_A \lambda_A q_R + h.c., \nonumber \\[2mm]
{\cal L}_{q\tilde q\chi^0_i} &=& iA^q_{\chi^0_i} \tilde q^\dagger_L \bar
\chi^0_i q_L + iB^q_{\chi^0_i} \tilde q^\dagger_R \bar\chi^0_i q_R + h.c.,
\nonumber \\[2mm]
{\cal L}_{q'\tilde q\chi^\pm_i} &=& iA^d_{\chi^\pm_i} \tilde u^\dagger_L
\bar\chi^-_i d_L + iA^u_{\chi^\pm_i} \tilde d^\dagger_L \bar\chi^+_i u_L +
h.c.,
\eea
where $g_s$ is the QCD coupling and $\lambda_A$ are the generators of the
colour $SU(3)$ group.  In the absence of top quark one can safely neglect the
small Yukawa couplings associated with the higgs sector, so that $A$ and
OB$B$ are simply the $SU(2) \times U(1)$ gauge couplings of left handed
(doublet) and right handed (singlet) quarks [5] respectively.   Moreover
we shall ignore the phase factors associated with these couplings, since the
interference terms between left and right handed squark exchanges are
negligible for final states involving only light quarks [22].  Thus we have
\bea
A^\mu_{\chi^0_i} &=& {g' \over 3\sqrt{2}} N_{i1} + {g \over \sqrt{2}}
N_{i2} \nonumber \\[2mm]
A^d_{\chi^0_i} &=& {g' \over 3\sqrt{2}} N_{i1} - {g \over \sqrt{2}} N_{i2}
\nonumber \\[2mm]
B^u_{\chi^0_i} &=& {4 \over 3} {g' \over \sqrt{2}} N_{i1} \nonumber
\\[2mm] B^d_{\chi^0_i} &=& -{2 \over 3} {g' \over \sqrt{2}} N_{i1}
\nonumber \\[2mm]
A^d_{\chi^\pm_i} &\equiv& A_{\chi^\pm_{iR}} = g U_{i1} \nonumber \\[2mm]
A^u_{\chi^\pm_i} &\equiv& A_{\chi^\pm_{iL}} = g V_{i1}
\eea
where $g$ and $g'$ are the standard $SU(2)$ and $U(1)$ couplings and $u,d$
stand for up and down member of any quark generation.  In terms of these
couplings we have the following spin-averaged squared matrix elements for
(21) and (22).
\be
\bar M^2_{\tilde g \rightarrow q\bar q \chi^0_i} =
\left(A^{q^2}_{\chi^0_i} + B^{q^2}_{\chi^0_i}\right) \left[{(\tilde g
\cdot \bar q) (\chi^0_i \cdot q) \over (m^2_{\tilde q} - (\tilde g - \bar
q)^2)^2} + {(\tilde g \cdot q) (\chi^0_i \cdot \bar q) \over (m^2_{\tilde
q} - (\tilde g - q)^2)^2}\right]
\ee
\be
\bar M^2_{\tilde g \rightarrow q' \bar q \chi^\pm_i} =
\left(A^{u^2}_{\chi^\pm_i} + A^{d^2}_{\chi^\pm_i}\right) \left[{(\tilde g
\cdot \bar q) (\chi^\pm_i \cdot q') \over (m^2_{\tilde q} - (\tilde g -
\bar q)^2)^2} + {(\tilde g \cdot q') (\chi^\pm_i \cdot \bar q) \over
(m^2_{\tilde q} - (\tilde g - q')^2)^2}\right]
\ee
where particle indices have been used for their 4-momenta and we have
ignored a common multiplicative constant involving the QCD coupling and
colour factor, since it is not relevant for our calculation [22].  Note that in
the limit $m_{\tilde g} \gg m_{\chi^0_i}, m_{\chi^\pm_i}$ the various
partial widths are propertional to the respective factors infront of the
square bracket.  Thus one gets the following branching fractions as a
simple first approximation.
\be
B_{\tilde g \rightarrow \chi^0_i} \simeq {{5 \over 2} N^2_{i2} + {49 \over
18} \tan^2\theta_W N^2_{i1} - {2 \over 3} \tan\theta_W N_{i1} N_{i2} \over
{13 \over 2} + {49 \over 18} \tan^2\theta_W} \simeq {2.5 N^2_{i2} + 0.83
N^2_{i1} - 0.37 N_{i1} N_{i2} \over 7.33}
\ee
\be
B_{\tilde g \rightarrow \chi^\pm_i} \simeq {2 U^2_{i1} + 2V^2_{i1} \over
{13 \over 2} + {49 \over 18} \tan^2\theta_W} \simeq {2U^2_{i1} + 2V^2_{i1}
\over 7.33}.
\ee
They show that the $SU(2)$ gauge interaction dominates over the $U(1)$ in
gluino decay.  Consequently the charginos account for a little over 50\%
of gluino decay and the neutralinos a little under 50\%, of which only
12\% goes into the $\tilde B$ dominated neutralino and the remainder into
the $\tilde W^3$ dominated one.  The gluino branching fractions into the
different neutralino and chargino states, resulting from (25) and (26),
are shown in Table I for the allowed range of the parameter space.  At
$m_{\tilde g} = 1000$ GeV they are shown only for $\tan\beta = 2$, since
in this case the loop induced decay processes are expected to become
significant for $\tan\beta = 10$ [2].  The branching fractions have been
obtained with a common squark mass
\be
m_{\tilde q} = m_{\tilde g} + 200 ~{\rm GeV},
\ee
but are insensitive to the choice of this parameter.  In fact one can
easily check that they are reasonably close to those obtained from the
approximate formulae (27) and (28).  One should note the following
systematic features which will be useful for our subsequent analysis.

\begin{enumerate}

\item[{1)}] The branching fraction for direct gluino decay into the LSP
has its maximum value for $\tilde B$ dominated LSP, i.e.
\be
B_{\tilde g \rightarrow \chi (\tilde B)} = .15 - .20.
\ee
It holds for most of the parameter space at $m_{\tilde g} = 300$ GeV and
for $\mu \simeq \pm 4 ~m_W$ at the higher values of gluino mass.

\item[{2)}] The largest branching fraction for gluino decay is into the
$\tilde W$ dominated chargino state, i.e.
\be
B_{\tilde g \rightarrow \chi^\pm_i (\tilde W)} \simeq 0.5.
\ee
It generally corresponds to the lighter (heavier) chargino state for $\mu
\simeq \pm 4 ~m_W ~(-m_W)$; but it has a substantial admixture of the
heavier (lighter) one at $m_{\tilde g} = 1000~(300)$ GeV as discussed
earlier.

\item[{3)}] The second largest branching fraction is into the
corresponding $\tilde W$ dominated neutralino, i.e.
\be
B_{\tilde g \rightarrow \chi^0_i (\tilde W)} \simeq 0.3.
\ee
It generally corresponds to the second lightest (heaviest) neutralino
state for $\mu \simeq \pm 4 ~m_W ~(-m_W)$, but again with the same caveat
as above.

\item[{4)}] Thus the $\tilde W$ dominated chargino and neutralino states
together account for $\sim 80\%$ of gluino decay.  Note that these two
states have nearly degenerate mass
\be
m_{\chi^\pm_i (\tilde W)} \simeq m_{\chi^0_i (\tilde W)}
\ee
throughout the parameter space; and this common mass is also roughly equal
to 1/3rd of the gluino mass, as expected from (10).  The first equality
implies very similar kinematics for the two major decay processes (31) and
(32); while the second implies that the kinematics is mainly determined by
the gluino mass and not by $\mu$ or $\tan\beta$.  As a result one gets a
fairly simple and robust signature as we shall see later.

\item[{5)}] The major decay modes of the $\tilde W$ dominated chargino and
neutralino states are
\be
\chi^\pm_i (\tilde W) {\buildrel W \over \longrightarrow} \chi^0_{1,2,3}
q' \bar q (\ell\nu),
\ee
\be
\chi^0_i (\tilde W) {\buildrel Z \over \longrightarrow} \chi^0_{1,2,3}
q\bar q (\ell\bar\ell).
\ee
Simple phase space considerations ensure that the decay into the LSP
$\chi$ $(\equiv \chi^0_1)$ dominates over most of the parameter space for
$m_{\tilde g} = 300$ GeV and at $\mu \simeq \pm 4~m_W$ for heavier
gluinos.  Note that the leptonic branching fractions of (34) and (35) are
about 20\% and 6\% respectively, where $\ell$ includes both $e$ and $\mu$.
These are the primary sources of isolated leptons from cascade decay [23].

\item[{6)}] Finally since all the branching fractions and kinematics of
gluino decay discussed above are very similar for $\mu = \pm 4~m_W$, we
shall present the resulting signals for only one of these two values.
\end{enumerate}

\newpage

\noindent IV. \underbar{\bf LSP Decay in $R\!\!\!/$ SUSY Models} \\
\bigskip

\nobreak
We shall concentrate on explicit $R$ parity violation [8-10], where
the LSP decay arises from one of the following $R\!\!\!/$ Yukawa
interaction terms in the Lagrangian.
\be
{\cal L}_{R\!\!\!/} = \lambda_{ijk} \ell_i \tilde \ell_j \bar e_k +
\lambda'_{ijk} \ell_i \tilde q_j \bar d_k + \lambda^{\prime\prime}_{ijk}
\bar d_i \tilde{\bar d}_j \bar u_k
\ee
plus analogous terms from the permutation of the supertwiddle.  Here
$\ell$ and $\bar e$ $(q$ and $\bar u,\bar d)$ denote the left handed
lepton doublet and antilepton singlet (quark doublet and antiquark
singlet) and $i,j,k$ are the generation indices.  The first two terms
correspond to $L\!\!\!/$ and the third one to $B\!\!\!/$ Yukawa
interaction.  While proton stability prohibit simulataneous presence of
both these interactions at any level of phenomenological significance,
either one of them could be present at a significant level.  Thus one has
two types of models, corresponding to $L\!\!\!/$ and $B\!\!\!/$.  In the
$B\!\!\!/$ model
\be
\chi \rightarrow d_i d_j u_k,
\ee
so that the only leptons in the signal are those coming from the cascade
decay.  Moreover the decay quarks from (37) lead to a stronger isolation
cut for these leptons, so that one expects a weaker signal in this case
compared to the $R$ conserving SUSY model.  On the other hand the
$L\!\!\!/$ model implies additional leptons from the LSP decay, resulting
in a much stronger signal as we see below.

In analogy with the standard Yukawa coupling of quarks and leptons to the
higgs boson one expects a hierarchical structure for these Yukawa
couplings as well.  Thus the dominant LSP decay process is
\be
\chi \rightarrow \ell_i q_j \bar d_k (e_i u_j \bar d_k + \nu_i d_j \bar d_k)
\ee
or
\be
\chi \rightarrow \ell_i \ell_j \bar e_k (e_i \nu_j \bar e_k + \nu_i e_j
\bar e_k)
\ee
depending on whether the dominant $L\!\!\!/$ Yukawa coupling is one of the
$\lambda'_{ijk}$ or $\lambda_{ijk}$ couplings (in the latter case particle
identity requires $i \not= j$).  The spin averaged and squared matrix
element for the decay process (39) is
\bea
\bar M^2_{\chi \rightarrow e_i \nu_j \bar e_k} &=& \bar M^2_{\chi
\rightarrow \nu_i e_j \bar e_k} = {A^{e^2}_\chi (\chi \cdot e) (\nu \cdot
\bar e) \over D^2_{\tilde e}} + {A^{\nu^2}_\chi (\chi \cdot \nu) (e \cdot
\bar e) \over D^2_{\tilde \nu}} \nonumber \\[2mm]
& & + {B^{e^2}_\chi (\chi \cdot \bar e) (e \cdot \nu) \over
D^2_{\tilde{\bar e}}} - {A^e_\chi A^\nu_\chi G(\chi,e,\bar e,\nu) \over
D_{\tilde e} D_{\tilde \nu}} + {A^e_\chi B^e_\chi G(\chi,e,\nu,\bar e)
\over D_{\tilde e} D_{\tilde{\bar e}}} \nonumber \\[2mm]
& & ~~~~~~~~~~~~~~~~~~~~~~~~~ + {A^\nu_\chi B^e_\chi G(\chi,\nu,e,\bar e) \over
D_{\tilde \nu} D_{\tilde{\bar e}}}, \nonumber
\eea
\be
D_{\tilde e} = m^2_{\tilde e} - (\chi - e)^2, ~G(\chi,e,\nu,\bar e) =
(\chi \cdot e) (\nu \cdot \bar e) - (\chi \cdot \nu) (e \cdot \bar e) +
(\chi \cdot \bar e) (\nu \cdot e),
\ee
and
\be
A^e_\chi = {-g' \over \sqrt{2}} N_{11} - {g \over \sqrt{2}} N_{12},
{}~A^\nu_\chi = {-g' \over \sqrt{2}} N_{11} + {g \over \sqrt{2}} N_{12},
{}~B^e_\chi = {-2g' \over \sqrt{2}} N_{11},
\ee
where we have dropped a common multiplicative factor involving the
$\lambda$ coupling, since it is not relevant for our calculation.
Moreover, one can factor out a common denominator assuming a common
slepton mass
\be
m_{\tilde e} = m_{\tilde \nu} \gg m_\chi.
\ee
We shall be working in this limit.

For simplicity we shall present the like sign dilepton signal for the
$L\!\!\!/$ SUSY model assuming the leading Yukawa coupling to be
$\lambda_{123}$.  This corresponds to a lepton (i.e. $e$ and $\mu$)
multiplicity of 1 for each LSP decay.  The corresponding lepton
multiplicities for the choices of different $\lambda$'s as the leading
$L\!\!\!/$ Yukawa coupling is listed in Table II.  The corresponding LSD
signals can be simply obtained by scaling the present signal by the
squares of the lepton multiplicities; for the lepton spectrum is
insensitive to the detailed structure of the squared matrix element.

If the leading $L\!\!\!/$ Yukawa coupling is a $\lambda'$ coupling, then
the relevant LSP decay is (38).  The corresponding squared matrix elements
are easily obtained from (40) by
obvious substitutions (see eq. 34 of [24]).  One should note however that
in this case the squared matrix elements for $\chi \rightarrow e u \bar d$
and $\nu d \bar d$ are not identical.  Consequently the lepton
multiplicity for this LSP decay is 1/2 only if the LSP is a pure $\tilde
B$ or $\tilde W$, but not for a general composition of LSP.  In the latter
case it can be calculated from the relative rates of the two decay
processes, i.e.
\be
{\Gamma_{\chi \rightarrow e u \bar d} \over \Gamma_{\chi \rightarrow \nu d
\bar d}} = {A^{e^2}_\chi + A^{u^2}_\chi + B^{d^2}_\chi - A^e_\chi A^u_\chi
+ A^e_\chi B^d_\chi + A^u_\chi B^d_\chi \over A^{\nu^2}_\chi +
A^{d^2}_\chi + B^{d^2}_\chi - A^\nu_\chi A^d_\chi + A^\nu_\chi B^d_\chi +
A^d_\chi B^d_\chi}
\ee
assuming $m_{\tilde e} \simeq m_{\tilde q} \gg m_\chi$.  The resulting
lepton multiplicities for the parameter values of our interest are shown
in Table III.  The corresponding LSD signals can again be obtained by
scaling the ones presented here by the squares of these multiplicities.
Although the lepton isolation cut is somewhat stronger in this case it
would not
degrade the signal substantially.  One should note that there would be no
$e$ or $\mu$ in LSP decay if the leading $L\!\!\!/$ coupling is a
$\lambda'_{3jk}$ or $\lambda'_{i3k}$ [10].  The first decay proceeds
through $\tau$ or $\nu_\tau$ emission and the second through $\nu_i$ only
due to the large top quark mass.

Finally it should be noted that we have conservatively assumed the leading
$R\!\!\!/$ Yukawa coupling to be $\ll 1$, so that the pair production of
superparticles and their decays into LSP are not affected [10].
\bigskip

\noindent V. \underbar{\bf The LSD Signal and Background} \\
\bigskip

\nobreak
The gluino pair production cross-section has been calculated for the
leading order QCD process [25]
\be
gg \rightarrow \tilde g\tilde g
\ee
using the gluon structure functions of [26] with a QCD scale $Q =
2m_{\tilde g}$.  Each of the gluinos is assumed to decay into the LSP via
the cascade decay processes discussed above.  The LSP escapes undetected
in the $R$ conserving SUSY model, while it decays in to a baryonic
(leptonic) channel in the $B\!\!\!/ ~(L\!\!\!/)$ violating models [8-10].
The resulting like sign dilepton signals have been calculated for the LHC
energy using a parton level Monte Carlo program.

The isolated LSD signals are shown in Figs. 1-5 along with the SM
background against the $p_T$ of the 2nd (softer) lepton, with the
isolation cut of eq. (6) and a rapidity cut of $|\eta| < 3$ on both the
leptons.  The SM background, arising from the $b\bar b$ production
(crosses) and $t\bar t$ production (histogram), were calculated in [12]
assuming $m_t = 150$ GeV [27].  The former dominates in the small $p_T$
region, while the latter dominates in the large $p_T$ region of our
interest.  But both are seen to become negligible for $p_{T2} \geq 60$
GeV.  The dominant background in this region is expected to arise from the
misidentification of one of the lepton charges in
\be
t\bar t \rightarrow \ell^+\ell^- X.
\ee
We have calculated the resulting fake LSD background, assuming it to be
about 1\% of the above cross-section, i.e. a misidentification of one of
the lepton charges at the 1/2\% level.  To avoid overcrowding, this
background has been shown only in Figs. 4 and 5.

Fig. 1 shows the isolated LSD signals for a 300 GeV gluino at $\mu =
4~m_W$ and both values of $\tan\beta$.  A brief discussion of the signal
curves is in order.

\begin{enumerate}

\item[{1)}] $R$ conserving Model: The main source of the signal in this
case is the squence
\be
\tilde g {\buildrel 0.5 \over \longrightarrow} \chi^\pm_i (\tilde W)
{\buildrel 0.2 \over \longrightarrow} \chi \ell^\pm \nu,
\ee
which has a leptonic branching fraction of 0.10.  The corresponding
branching fraction for the second largest source
\be
\tilde g {\buildrel 0.3 \over \longrightarrow} \chi^0_i (\tilde W)
{\buildrel 0.06 \over \longrightarrow} \chi \ell^+ \ell^-
\ee
is effectively 0.036.  Moreover the degeneracy relation (33) along with
$m_W \simeq m_Z$ imply very similar kinematic distributions for the two
final states.  Thus one can simply take account of the second source by
increasing the branching fraction of the first by 36\%.  We have followed
this prescription in obtaining the LSD signal.  Following this procedure
we have calculated the dilepton cross-section assuming the decay chain
(46) for both the gluinos, with $\chi^\pm_i (\tilde W) = \chi^\pm_1$.  The
resulting dilepton branching fraction is $\simeq 2\%$.  Deviding it by a
factor of 2 gives the final LSD signal, shown as the dot-dashed lines.

\item[{2)}] $B\!\!\!/$ Model: The source of the dileptons in this case is
the same as above.  However, the quarks coming from the LSP decay (37) are
included in the isolation cut of the leptons.  This results in a
substantial depletion of the LSD signal as shown by the short dashed
lines.

\item[{3)}] $L\!\!\!/$ Model: The main source of the LSD signal in this
case are the leptons from the LSP decay.  The hardest component
corresponds to the direct decay of each gluino into the LSP, i.e.
\be
\tilde g {\buildrel .17-.20 \over \longrightarrow} \chi(\tilde B)
\longrightarrow \ell.
\ee
Thus the dilepton branching fraction is $\simeq 3-4\%$, i.e. roughly
similar to the $R$
conserving case.  The resulting LSD signal, shown by the dotted lines, is
similar to the later in both shape and size.  However, the largest
component comes from the decay sequence
\be
\tilde g {\buildrel 0.8 \over \longrightarrow} \chi^\pm_i (\tilde W),
{}~\chi^0_i (\tilde W) \rightarrow \chi \rightarrow \ell
\ee
for each gluino.  This can be combined with the cross-term between (48)
and (49), since the 2nd (softer) lepton in either case comes from (49).
Hence the combined dilepton branching fraction is $\simeq 1$.  The
resulting LSD signal is shown by the solid lines.  Finally the cross-term
between the combined decay sequence (48) and (49) for one gluino and (46)
for the other has a dilepton branching fraction of $\simeq 0.2$, which is
shown by the long dashed line [28].  There is negligible double counting
in adding the above two components, since the probability of both the
decay leptons coming from the same gluino to populate the large $p_T$
region of interest is negligible.  As mentioned earlier, the $L\!\!\!/$
signals have been calculated for the LSP decay mode $\chi \rightarrow \ell
\nu \tau$ having a leptonic multiplicity of 1.  The corresponding signals
for the other $L\!\!\!/$ decays of (38) and (39) can be obtained by
multiplying them with the respective leptonic multiplicities of each LSP
decay, shown in Tables II and III.
\end{enumerate}

As one sees from Fig. 1, the
$L\!\!\!/$ LSD signal is clearly large compared to the SM background at
large $p_T$.  It is also larger than the fake LSD background shown in Fig.
5.  The $R$ conserving signal is larger than the first but comparable to
the second.  Nonetheless it can be easily recognised by the large
missing-$p_T$ carried by the $\chi$ and $\nu$ of (46).  This is shown in
Fig. 6.  The $B\!\!\!/$ signal is somewhat larger than the SM background
but smaller than that coming from the fake LSD by a factor of $\sim 5$.
Thus identifying this signal would require identification of lepton charge
to a 0.1\% accuracy.
Fig. 2 shows the corresponding signals for $\mu =
-m_W$.   In this case the gluino has significant branching fractions into
both $\chi^\pm_1$ and $\chi^\pm_2$; but the former is still the larger
one.  Moreover if either of the gluinos decays via $\chi^\pm_1$, the
softer lepton $p_T$ distribution would correspond to this decay mode.
Therefore it is reasonable to approximate $\chi^\pm_i (\tilde W)$ by
$\chi^\pm_1$ as in the previous case.  Thus the decay sequences and
branching fractions are identical to (46-49), except for a marginal
reduction of the dilepton branching fraction from (48) from 3 to 2\% at
$\tan\beta = 10$.   A comparison of the signal curves with those of
Fig. 1 shows that all of them are qualitatively similar.  Thus the 300 GeV
gluino signals are seen to be fairly insensitive to the choice of $\mu$ as
well as $\tan\beta$.

Fig. 3 shows the LSD signals for a 600 GeV gluino at $\mu = 4~m_W$.  Again
in this case the LSP is dominated by $\tilde B$; and the second lightest
neutralino and the lighter chargino are dominated by $\tilde W$.
Therefore we can use the gluino decays of (46-49) with the same branching
fractions; the dilepton branching fraction from (48) is 3\% for both
values of $\tan\beta$.  The resulting LSD signals are of course smaller
and harder
than the previous case.  Nonetheless the $L\!\!\!/$ and $R$ conserving
signals are confortably above the SM background for $p_{T2} \gsim 60$ GeV,
while the $B\!\!\!/$ signal is comparable to this background.  Moreover
the $L\!\!\!/$ violating signal is larger than the fake LSD background as
well.  Although the $R$ conserving signal is smaller than this
background, it can again be distinguished by the large missing-$p_T$
accompanying this signal (Fig. 6).  But it would be difficult to identify
the $B\!\!\!/$ signal unless the fake LSD background can be further
suppressed by an order of magnitude.  Fig. 4 presents the corresponding
LSD signals for $\mu = -m_W$.  Here the gaugino dominated states are the
heavier chargino and neutralinos.  Consequently the direct decay of gluino
into LSP (48) is negligible, while the cascade decays (46) and (47)
proceed via $\chi^\pm_2$ and $\chi^0_4$.  The signals of Fig. 4 have been
obtained with this substitution.  One noticable change is the enhancement
of the $B\!\!\!/$ signal; the higher mass of $\chi^\pm_2 ~(\chi^0_4)$
ensures the isolation of the decay lepton even in the presence of (37).
It should be added here that the $\chi^\pm_2 ~(\chi^0_4)$ decay has
significant branching fraction into $\chi^0_2$ [5], which has not been
taken into account in these curves.  However, we have checked its effect
on the $R$ conserving LSD signal by replacing $\chi$ by $\chi^0_2$ in
(46).  It has negligible effect on the lepton momentum spectrum and hence
the resulting LSD signal, since $m_{\chi^0_2} - m_\chi \ll m_{\chi^\pm_2}
- m_{\chi^0_2}$.  It may be noted here that this mass difference is also
small compared to $m_{\chi^0_2}$, particularly for $\tan\beta = 2$.
Consequently $\chi$ should carry a large part of the $\chi^0_2$ momentum
in the cascade decay $\chi^\pm_2 \rightarrow \chi^0_2 \rightarrow
\chi$; and hence the resulting $L\!\!\!/$ LSD signal should not be
substantially degraded.  However, we have not checked this quantitatively
since this signal is any way quite large.  Finally, a comparison of the
signal curves of Figs. 3 and 4 shows that they are quite similar for the
$L\!\!\!/$ as well as the $R$ conserving case.  The reason of course is
that the masses of the respective $\chi^\pm_i (\tilde W)$ states are
qualitatively similar, as remarked before.

Since the LSD signals for a 1000 GeV gluino are less promising, we have
presented them in Fig. 5 for only one set of parameters, i.e. $\mu =
4~m_W$ and $\tan\beta = 2$.  It also shows the SM as well as the fake LSD
background.  The $L\!\!\!/$ signal is larger than the first at large $p_T$
but somewhat below the second.  The latter can be reduced below the signal
if one can identify lepton charge to within 0.2\% accuracy.  It should
also be possible to separate the two via the accompanying ${p\!\!\!/}_T$
distribution.  The size of the $R$ conserving LSD signal is much too low,
while the $B\!\!\!/$ signal lies below the scale of this figure.  One
hopes the size of the $R$ conserving LSD signal to become viable at the
SSC energy or the high luminosity option of LHC.  Although the signal to
background ratio is not expected to improve, the two can be separated via
the accompanying ${p\!\!\!/}_T$ distribution (Fig. 6).  Of course the
canonical missing-$p_T$ signal of a 1000 GeV gluino is expected to be
observable even at the low luminosity option of LHC [2].

Let us conclude this section by looking at the fate of the LSD signal if
squarks are lighter than gluinos.  In this case the superparticle
production would be dominated by pair production of squarks, in which
singlet and doublet pairs occur with equal probability.  Since the singlet
squarks do not couple to $\tilde W$, the major decay mode is through
$\tilde B$ dominated neutralino.  Consequently the direct decay into
$\chi$ is expected to dominate over a large part of the parameter space
[5].  The doublet squark decays are very similar to the gluino decays,
except that the decay of the chargino pair would always lead to opposite
sign dileptons.  The largest source of LSD is the decay of one squark via
$\chi^\pm_i (\tilde W)$ (46) and the other via $\chi^0_i (\tilde W)$ (47).
The end result is a degradation of the $R$ conserving LSD signal by a
factor of $\sim 5$, while the $L\!\!\!/$ signal is enhanced compared to
the gluino case.
\bigskip

\noindent VI. \underbar{\bf Summary} \\
\bigskip

\nobreak
We have investigated the isolated LSD signal for gluino production at the
LHC energy in the $R$ conserving as well as $L$ and $B$ violating SUSY
models.  The signals are investigated for representative values of gluino
mass, $\mu$ and $\tan\beta$ assuming MSSM, against the standard model
background as well as the fake LSD background coming from the
misidentification of lepton charge.  The main results are listed below.

\begin{enumerate}

\item[{1)}] The signals are fairly insensitive to the choice of
$\tan\beta$ as well as the $\mu$ parameter.  Thus for a given gluino mass
one has fairly robust signals, which should hold for at least the bulk of
the $\tan\beta$ and $\mu$ parameter space.

\item[{2)}] For the $L\!\!\!/$ SUSY model, the LSD signal is larger than
both the SM and the fake LSD background for gluino masses of 300 and 600
GeV.  Although the $R$ conserving signal is somewhat lower than the latter
background, it can be identified via the large missing-$p_T$ accompanying
the LSD.  The signal is less promising, however, for the $B\!\!\!/$ case.

\item[{3)}] For a gluino mass of 1000 GeV, the $L\!\!\!/$ LSD signal
should still be viable.  However the size of the $R$ conserving LSD signal
is too small in this case, at least for the low luminosity option of LHC.

\item[{4)}] In going from pair production of gluinos to that of squarks
one expects a substantial degradition of the LSD signal for the $R$
conserving model while it is expected to be enhanced for the $L\!\!\!/$
model.

\end{enumerate}
\bigskip

\noindent \underbar{\bf Acknowledgement}: We gratefully acknowledge
discussions with Amitava Datta, Manual Drees, Rohini Godbole, Graham Ross,
Probir Roy and Xerxes Tata.  One of us (MG) acknowledges the hospitality of
the Theory Group of TIFR during the course of this work.

\newpage

\noindent \underbar{\bf References} \\
\bigskip

\begin{enumerate}

\item[{1.}] For a review see e.g. H. Haber and G. Kane, Phys. Rep. 117, 75
(1985).

\item[{2.}] Report of the Supersymmetry Working Group (C. Albajar et al.),
Proc. of ECFA-LHC Workshop, Vol. II, p. 606-683, CERN 90-10 (1990).

\item[{3.}] The missing-$p_T$ signature is used for superparticle search
at the LEP $e^+e^-$ collider as well; but it plays a less crucial role
there, since most of their constraints on superparticle masses simply
follow from the measurement of $Z$ total width.  See e.g. ALEPH
collaboration, D. Decamp et al., Phys. Rep. 216, 253 (1992).

\item[{4.}] J. Ellis, J. Hagelin, D.V. Nanopoulos, K. Olive and M.
Srednicki, Nucl. Phys. B238, 453 (1984).

\item[{5.}] H. Baer, V. Barger, D. Karatas and X. Tata, Phys. Rev., D36,
96 (1987).

\item[{6.}] M. Barnett, J. Gunion and H. Haber, Proc. of 1988 Summer Study
on High Energy Physics, Snowmass, Colorado (World Scientific, 1989) p.
230; Proc. of 1990 Summer Study on High Energy Physics, Snowmass, Colorado
(World Scientific, 1992) p. 201.

\item[{7.}] H. Baer, X. Tata and J. Woodside, Phys. Rev. D41, 906 (1990).

\item[{8.}] S. Dawson, Nucl. Phys. B261, 297 (1985); S. Dimopoulos and
L.J. Hall, Phys. Lett. B207, 210 (1987); S. Dimiopoulos et al., Phys. Rev.
D41, 2099 (1990).

\item[{9.}] H. Dreiner and G.G. Ross, Nucl. Phys. B365, 597 (1991).

\item[{10.}] D.P. Roy, Phys. Lett. B283, 270 (1992).

\item[{11.}] CDF collaboration: F. Abe et al., Phys. Rev. Lett. 69, 3439
(1992).

\item[{12.}] N.K. Mondal and D.P. Roy, TIFR TH/93/23, submitted to the
Phys. Rev. D.

\item[{13.}] R.M. Godbole, S. Pakvasa and D.P. Roy, Phys. Rev. Lett. 50,
1539 (1983); see also V. Barger, A.D. Martin and R.J.N. Phillips, Phys.
Rev. D28, 145 (1990).

\item[{14.}] D.P. Roy, Phys. Lett. B196, 395 (1987).

\item[{15.}] F. Cavanna, D. Denegri and T. Rodrigo, Proc. of ECFA-LHC
Workshop, Vol. II, p. 329, CERN 90-10 (1990).

\item[{16.}] F.E. Paige and S.D. Protopopescu, ISAJET Program, BNL-38034
(1986).

\item[{17.}] J.F. Gunion and H. Haber, Nucl. Phys. B272, 1 (1986).

\item[{18.}] M. Guchait, Z. Phys. C57, 157 (1993); see also M.M.
Elkheishen, A.A. Shafik and A.A. Aboshousha, Phys. Rev. D45, 4345 (1992).

\item[{19.}] L. Roszkowski, Phys. Lett. B262, 59 (1991).

\item[{20.}] The $\epsilon$ parameter of [5] corresponds to $-\mu$.  We
thank Xerxes Tata for a clarification of this point.  The main difference
between our Table I and the corresponding Table of [5] lies in the range
of $\tan\beta$ covered.  Besides, the more up to date values of the SM
parameters used here make a nonnegligible difference to the result.

\item[{21.}] Only a gauge dominated neutralino is not seriously
constrained by the LEP data [3] since it does not couple to $Z$.

\item[{22.}] There are also interference terms between the left handed
squark and right handed antisquark exchanges (and vice versa) which vanish
in the limit $m_{\chi^0_i} ~(m_{\chi^\pm_i}) \ll m_{\tilde g}$.  We shall
neglect them since the gluino decays mainly into the gauge dominated
chargino/neutralinos, for which this mass inequality is reasonably
satisfied by eqs. (10) and (11).  However we have checked that the gluino
branching fractions evaluated by retaining these interference terms [5]
agree with those shown in Table I to within 10\%.

\item[{23.}] There is one exception to this however, which occurs for
$m_{\tilde \ell}$ $< m_{\chi^\pm_i (\tilde W)}$ $< m_\chi + m_W$.  In this
case the chargino decays dominantly into a lepton via the slepton, so that
the resulting LSD signal is comparable in size to the canonical
missing-$p_T$ signal.  This case has been recently investigated by R.M.
Barnett, J.F. Gunion and H.E. Haber, LBL-34106 (1993); and H. Baer, C. Kao
and X. Tata, FSU-HEP-930527 (1993).

\item[{24.}] J. Butlerworth and H. Dreiner, OUNP-92-15, Nucl. Phys. B (to
be published).

\item[{25.}] P.R. Harrison and C.H. Llewellyn Smith, Nucl. Phys. B213, 223
(1983) and B223, 542 (E) (1983); E. Reya and D.P. Roy, Phys. Rev. D32, 645
(1985).

\item[{26.}] M. Gluck, E. Hoffman and E. Reya, Z. Phys. C13, 119 (1982).

\item[{27.}] These LSD background are 8 time larger than the ones shown in
[12], since they include both $e$ and $\mu$ and both the charge
combinations $++$ and $--$.

\item[{28.}] One could effectively include the decay sequence (47) along
with (46) by increasing the dilepton branching fraction and hence the
normalisation of the long dashed line by 36\%, though we have not done this.

\end{enumerate}

\newpage

\textwidth=16cm
\textheight=25.5cm
\topmargin=-3.5cm
\abovedisplayskip=3pt
\belowdisplayskip=3pt
\vbox{
\begin{center}
{\Large {\bf Table ~I(a)}} \\
Masses (in GeV) and compositions of the neutralino $\chi^0_i$ and chargino
$\chi^\pm_i$ \\ states along with the corresponding
gluino branching fractions for $m_{\tilde g} = 300$ GeV. \\
\end{center}
\[
\begin{tabular}{|c c c c c c c|}
\hline
\multicolumn{1}{|c} {$\mu$} &
\multicolumn{1}{c} {$m_{\chi^0_i}$} &
\multicolumn{1}{c} {$N_{ij}$} &
\multicolumn{1}{c} {$B_{\tilde g \rightarrow \chi^0_i}$} &
\multicolumn{1}{c} {$m_{\chi^\pm_i}$} &
\multicolumn{1}{c} {$V_{ij}/U_{ij}$} &
\multicolumn{1}{c|} {$B_{\tilde g \rightarrow \chi^\pm_i}$} \\
\hline
&& ~~~$\tan\beta=2$ &&&& \\ $4 m_W$ & 42.3 & -.94,.24,-.18,.10 & .20 &
77.7 & -.97,.23 & .47 \\ & 81.8 & -.29,.92,-.20,.13 & .32 & & -.93,.35
& \\ & -326.4 & -.03,.06,.69,.71 & 0 & 349.8 & .23,.97 & 0 \\ & 352.9
& -.13,.28,.66,-.68 & 0 & & .35,.93 & \\ $-4m_W$ & 53.6 &
-.98,-.11,.08,-.03 & .20 & 110.6 & .99,.06 & .47 \\ & 110.7 &
-.10,.97,.21,-.04 & .32 & & .95,.28 & \\ & 328.0 & -.05,.11,-.68,-.71
& 0 & 340.8 & -.06,.99 & 0 \\ & -341.9 & -.10,.16,-.68,.69 & 0 & &
.28,-.95 & \\ $-m_W$ & 54.9 & -.90,-.14,.38,.09 & .20 & 91.5 &
.81,-.57 & .30 \\ & 73.1 & .28,.42,.69,.50 & .11 & & .25,.96 & \\ &
-118.4 & -.23,.32,-.60,.68 & .04 & 146.2 & .57,.81 & .19 \\ & 141.3 &
-.20,.83,-.52,-.50 & .16 & & .96,-.25 & \\ $m_W$ & -.01 &
-.05,.46,-.65,.32 & & 16.6 & -.75,.65 & \\ & 63.3 & .82,.51,-.18,.17 &
& & -.55,.83 & \\ & -87.5 & -.13,.17,.60,.76 & & 171.6 & .65,.75 & \\
& 175.0 & -.23,.70,.41,-.53 & & & .83,.55 & \\ $0.1m_W$ & -6.2 &
-.04,.04,-.92,-.38 & & 29.8 & .67,-.74 & \\ & -53.9 &
-.34,.42,-.29,.77 & & & -.29,.95 &\\ & 59.8 & -.90,-.37,.09,-.16 & &
149.4 & .74,.67 & \\ & 150.9 & -.22,.82,.24,-.46 & & & .95,.29 &\\
\hline
&& ~~~$\tan\beta=10$ &&&& \\
$4m_W$ & 48.2 & -.98,.08,-.15,.04 & .17 & 89.8 & -.99,.13 & .52 \\
& 90.6 & .12,.95,-.24,.08 & .31 & & -.93,.36 & \\
& -332.5 & -.07,.01,.68,.71 & 0 & 346.9 & .13,.99 & 0 \\
& 344.3 & -.11,.24,.66,-.69 & 0 & & .36,.93 & \\
$-4m_W$ & 50.9 & -.99,-.01,.13,.01 & .18 & 97.9 & .99,-.07 & .50 \\
& 97.8 & .02,.96,.24,.05 & .31 & & .93,.34 & \\
& -336.3 & -.08,.14,-.68,.70 & 0 & 344.7 & .07,.99 & 0 \\
& 338.2 & -.09,.21,-.66,-.70 & 0 & & .34,-.93 & \\
$-m_W$ & 37.2 & -.62,.24,.69,.26 & .15 & 58.0 & .83,-.54 & .42 \\
& 65.0 & .71,.53,.35,.27 & .22 & & .38,.92 & \\
& -109.0 & -.21,.29,-.57,.73 & .04 & 162.4 & .54,.83 & .10 \\
& 157.5 & -.22,.75,-.24,-.56 & .08 & & .92,-.38 & \\
$m_W$ & 23.3 & -.54,.35,-.71,.26 & & 40.6 & -.82,.56 & \\
& 63.9 & .78,.52,-.25,.21 & & & -.43,.89 & \\
& -101.8 & -.19,.27,.56,.75 & & 167.6 & .56,.82 & \\
& 165.3 & -.22,.72,.31,-.56 & & & .89,.43 & \\
$0.1m_W$ & -.01 & -.06,.06,-.99,-.01 & & 3.2 & .74,-.66 & \\
& 57.6 & -.33,.42,.03,.84 & & & -.08,.99 & \\
& -59.7 & -.91,-.36,.04,-.18 & & 152.3 & .66,.74 & \\
& 149.7 & -.22,.82,.07,-.50 & & & .99,.08 & \\
\hline
\end{tabular}
\]
}

\newpage

\textwidth=16cm
\textheight=25.5cm
\topmargin=-3.5cm
\abovedisplayskip=3pt
\belowdisplayskip=3pt
\vbox{
\begin{center}
{\Large {\bf Table ~I(b)}} \\
Masses (in GeV) and compositions of the neutralino $\chi^0_i$ and chargino
$\chi^\pm_i$ states along with the corresponding gluino branching
fractions for $m_{\tilde g} = 600$ GeV. \\
\end{center}
\[
\begin{tabular}{|c c c c c c c|}
\hline
\multicolumn{1}{|c} {$\mu$} &
\multicolumn{1}{c} {$m_{\chi^0_i}$} &
\multicolumn{1}{c} {$N_{ij}$} &
\multicolumn{1}{c} {$B_{\tilde g \rightarrow \chi^0_i}$} &
\multicolumn{1}{c} {$m_{\chi^\pm_i}$} &
\multicolumn{1}{c} {$V_{ij}/U_{ij}$} &
\multicolumn{1}{c|} {$B_{\tilde g \rightarrow \chi^\pm_i}$} \\
\hline
&& ~~~$\tan\beta=2$ &&&& \\
$4m_W$ & 92.3 & -.96,.14,-.18,.12 & .18 & 164.5 & -.92,.38 & .49 \\
& 169.3 & .22,.89,-.29,.23 & .30 & & -.88,.47 & \\
& -326.1 & -.03,.05,.70,.70 & 0 & 362.4 & .38,.92 & .02 \\
& 365.8 & -.14,.41,.62,-.64 & .01 & & .47,.88 & \\
$-4m_W$ & 104.1 & -.99,-.05,.10,-.11 & .20 & 206.0 & .99,-.05 & .49 \\
& 205.5 & -.03,.96,.24,.05 & .30 & & .94,.33 & \\
& 330.6 & -.06,.20,-.67,-.71 & .01 & 340.5 & .05,.99 & .01 \\
& -339.0 & -.09,.13,-.69,.70 & 0 & & .33,-.94 &   \\
$-m_W$ & 74.0 & -.28,.11,.75,.57 & .03 & 95.5 & .43,-.89 & .08 \\
& -108.4 & -.19,.23,-.64,.70 & .04 & & -.04,.99 & \\
& 111.1 & -.93,-.18,-.10,-.28 & .19 & 225.0 & .89,.43 & .41 \\
& 224.6 & -.10,.94,.05,-.30 & .25 & & .99,.04 & \\
$m_W$ & 27.4 & -.43,.34,-.67,.48 & & 45.6 & -.51,.85 & \\
& -85.1 & -.09,.10,.64,.74 & & & -.35,.93 & \\
& 117.7 & -.88,-.32,.22,.25 & & 240.1 & .85,.51 & \\
& 241.1 & -.14,.87,.26,-.37 & & & .93,.35 & \\
$0.1m_W$ & -5.8 & -.05,.05,-.94,-.31 & & 15.8 & .44,-.89 & \\
& -35.5 & -.28,.30,.26,.86 & & & -.21,.97 & \\
& 112.1 & -.99,-.21,.11,.20 & & 230.2 & .89,.44 & \\
& 230.5 & -.12,.92,.16,-.31 & & & .97,.21 & \\
\hline
&& ~~~$\tan\beta=10$ &&&& \\
$4m_W$ & 97.6 & -.98,.06,-.16,.06 & .18 & 178.8 & -.95,.29 & .48 \\
& 180.2 & .12,.91,-.31,.19 & .31 & {} & -.88,.47 & \\
& -331.1 & -.06,.09,.69,.71 & 0 & 355.5 & .29,.95 & .01 \\
& 354.5 & -.12,.37,.62,-.67 & .01 &  & .47,.88 & \\
$-4m_W$ & 100.5 & -.98,.01,.14,.03 & .19 & 188.9 & .97,-.22 & .49 \\
& 189.0 & .06,.93,.31,.16 & .30 & & .89,.44 & \\
& 334.3 & -.07,.11,-.69,.70 & 0 & 350.3 & .22,.97 & .01 \\
& 346.0 & -.11,.33,-.63,-.68 & .01 &  & .44,-.89 & \\
$-m_W$ & 55.9 & -.38,.22,.75,.47 & .06 & 75.4 & .53,-.84 & .14 \\
& -100.8 & -.17,.21,-.61,.73 & .03 & & .14,.98 & \\
& 114.9 & -.89,-.26,-.19,-.29 & .18 & 232.5 & .84,.53 & .35 \\
& 231.3 & -.12,.91,-.09,-.37 & .23 & & .98,.14 & \\
$m_W$ & 44.8 & -.41,.27,-.74,.45 & & 63.2 & -.54,.83 & \\
& -95.2 & -.15,.18,.61,.74 & & &-.22,.97 & \\
& 116.3 & -.88,-.28,.21,-.28 & & 236.1 & .83,.54 & \\
& 235.4 & -.13,.89,.16,-.38 & & & .97,.22 & \\
$0.1m_W$ & -.03 & -.06,.06,-.99,.06 & & 1.4 & -.49,.86 & \\
& -40.2 & -.27,.29,.09,.90 & & & -.06,.99 & \\
& 111.9 & -.95,-.20,.04,-.22 & & 230.7 & .86,.49 & \\
& 229.9 & -.11,.92,.05,-.34 & & & .99,.06 & \\
\hline
\end{tabular}
\]
}

\newpage

\vbox{
\begin{center}
{\Large {\bf Table ~I(c)}} \\
Masses (in GeV) and compositions of the neutralino $\chi^0_i$ and chargino
$\chi^\pm_i$ states along with the corresponding gluino branching
fractions for $m_{\tilde g} = 1000$ GeV. \\
\end{center}
\[
\begin{tabular}{|c c c c c c c|}
\hline
\multicolumn{1}{|c} {$\mu$} &
\multicolumn{1}{c} {$m_{\chi^0_i}$} &
\multicolumn{1}{c} {$N_{ij}$} &
\multicolumn{1}{c} {$B_{\tilde g \rightarrow \chi^0_i}$} &
\multicolumn{1}{c} {$m_{\chi^\pm_i}$} &
\multicolumn{1}{c} {$V_{ij}/U_{ij}$} &
\multicolumn{1}{c|} {$B_{\tilde g \rightarrow \chi^\pm_i}$} \\
\hline
&& ~~~$\tan\beta=2$ &&&& \\
$4m_W$ & 156.5 & -.94,.12,-.22,.18 & .17 & 252.8 & -.71,.70 & .31 \\
& 262.6 & .28,.69,-.48,.44 & .19 & & -.65,.75 & \\
& -325.8 & -.03,.04,.70,.71 & 0 & 406.9 & .70,.71 & .20 \\
& 408.7 & -.12,.70,.47,-.51 & .12 & & .75,.65 & \\
$-4m_W$ & 170.9 & -.99,-.02,.11,.004 & .17 & 311.7 & .72,-.68 & .23 \\
& 307.1 & .06,.58,.61,.52 & .12 & & .55,.83 & \\
& -336.4 & -.08,.11,-.69,.70 & .01 & 363.8 & .68,.72 & .27 \\
& 360.5 & -.06,.80,-.34,-.47 & .19 & & .83,-.55 & \\
$-m_W$ & 77.5 & -.12,.08,.74,.64 & .01 & 92.3 & .27,-.96 & \\
& 101.0 & -.15,.17,-.66,.71 & .02 & & -.08,.99 & \\
& 176.2 & -.97,-.08,.01,-.18 &.15 & 349.7 & .96,.27 & \\
& 349.4 & -.05,.97,.06,-.19 &.30 & & .99,.08 & \\
$m_W$ & 47.7 & -.03,.24,-.70,.59 &  & 61.1 & -.32,.94 & \\
& -83.7 & -.01,.01,.66,.73 &  & & -.21,.97 & \\
& 181.7 & -.94,-.14,.17,-.22 &  & 356.5 & .94,.32 & \\
& 356.4 & -.06,.95,.15,-.23 & & & .97,.21 & \\
$0.1m_W$ & -5.1 & .90,-.37,-.10,.17 & & 7.3 & .29,-.95 & \\
& 22.9 & -.42,-.87,-.11,.20 & & & -.14,.98 & \\
& 177.9 & .003,-.08,.97,.20 & & 352.5 & .95,.29 & \\
& 352.3 & .08,-.28,.16,-.94 & & & .98,.14 & \\
\hline
&& ~~~$\tan\beta=10$ &&&& \\
$4m_W$ & 162.5 & -.96,.06,-.20,.11 & & 269.9 & -.73,.67 & \\
& 274.1 & .21,.68,-.52,.45 &  & {} & -.62,.78 & \\
& -329.9 & -.056,.076,.69,.71 & & 395.8 & .67,.73 &  \\
& 395.4 & -.10,.72,.44,-.52 & & & .78,.62 & \\
$-4m_W$ & 166.1 & -.98,.03,.18,.07 &  & 283.3 & .74,-.67 &  \\
& 283.9 & -.06,.92,-.69,.70 &  & & .60,.79 & \\
& -332.5 & -.07,.09,-.07,.70 & & 386.3 & .67,.74 & \\
& 384.6 & -.09,.73,-.42,-.51 &  &  & .79,-.60 & \\
$-m_W$ & 66.5 & -.22,.15,.75,.59 &  & 80.3 & .33,-.94 & \\
& -95.0 & -.13,.15,-.64,.73 & & & .05,.99 & \\
& 178.6 & -.96,-.11,-.08,-.22 & & 352.7 & .94,.33 & \\
& 352.1 & -.05,.97,-.03,-.23 & & & .99,-.05 & \\
$m_W$ & 59.2 & -.26,.18,-.74,.57 & & 72.6 & -.34,.93 & \\
& -90.9 & -.11,.12,.64,.74 & & &-.11,.99 & \\
& 179.9 & -.95,-.12,.12,-.23 & & 354.3 & .93,.34 & \\
& 353.8 & -.06,.96,.08,-.24 & & & .99,.11 & \\
$0.1m_W$ & .05 & -.06,.06,-.98,.15 & & 4.0 & -.32,.94 & \\
& -28.2 & -.20,.20,.18,.94 & & & -.04,.99 & \\
& 177.7 & -.97,-.10,.03,-.20 & & 352.6 & .94,.32 & \\
& 352.1 & -.05,.97,.03,.22 & & & .99,.04 & \\
\hline
\end{tabular}
\]
}

\newpage

\vbox{
\begin{center}
{\Large {\bf Table ~II}} \\
Multiplicity of lepton $(e,\mu)$ per LSP decay for leading $R\!\!\!/$
Yukawa coupling $\lambda_{ijk}$, valid for any gauge composition of LSP. \\
\end{center}
\[
\begin{tabular}{|c|c|c|c|c|}
\hline
\multicolumn{1}{|c} {Leading} &
\multicolumn{1}{|c|} {$\lambda_{123}$} &
\multicolumn{1}{|c|} {$\lambda_{1(2)33}$} &
\multicolumn{1}{|c|} {$\lambda_{1(2)31(2)}$} &
\multicolumn{1}{c|} {$\lambda_{121(2)}$} \\
coupling &&&& \\
\hline
Multiplicity & 1 & 1/2 & 3/2 & 2 \\
of $e(\mu)$ &&&& \\
\hline
\end{tabular}
\]
}
\bigskip
\bigskip
\bigskip

\vbox{
\begin{center}
{\Large {\bf Table ~III}} \\
Multiplicity of lepton $(e,\mu)$ per LSP decay for leading $R\!\!\!/$
Yukawa coupling $\lambda'_{ijk} (i,j \neq 3)$ and different gauge
composition of LSP. \\
\end{center}
\[
\begin{tabular}{|c c c|}
\hline
\multicolumn{1}{|c} {$m_{\tilde g}$ } &
\multicolumn{1}{c} {300 GeV} &
\multicolumn{1}{c|} {600 GeV} \\
$\tan\beta$ & 2 ~~~~ 10 & 2 ~~~~ 10 \\
\hline
$\mu = -4m_W$ & .67 ~~ .52 & .58 ~~ .48 \\
$\mu = 4m_W$ & .20 ~~ .38 & .30 ~~ .41 \\
$\mu = -m_W$ & .72 ~~ .13 & .13 ~~ .13 \\
\hline
\end{tabular}
\]
}

\newpage

\begin{center}
{\bf Figure Captions} \\
\end{center}
\bigskip

\begin{enumerate}

\item[{\rm Fig.~ 1.}] The LSD signals for 300 GeV gluino production at LHC
at $\mu = 4~m_W$ and $\tan\beta = 2,10$ shown against the $p_T$ of the 2nd
(softer) lepton.  The dot-dashed and short dashed lines represent the
signals in the $R$ conserving and $B$ violating SUSY models, while the
solid, long dashed and dotted lines are three components of the signal in
the $L$ violating model.  The SM background from $t\bar t$ via sequencial
decay and from $b\bar b$ via mixing are shown by the histogram and the
crosses respectively.

\item[{\rm Fig.~2.}] The LSD signals for 300 GeV gluino production at $\mu
= -m_W$ shown along with the SM background.  The conventions are the same
as in Fig. 1.

\item[{\rm Fig.~3.}] The LSD signal for 600 GeV gluino production at $\mu
= 4~m_W$ shown along with the SM background.  The conventions are the same
as in Fig. 1.

\item[{\rm Fig.~4.}] The LSD signal for 600 GeV gluino production at $\mu
= -m_W$ shown along with the SM background, with the same convention as in
Fig. 1.  The circles denote the fake LSD background from misidentification
of one of the lepton charges.

\item[{\rm Fig.~5.}] The LSD signal for 1000 GeV gluino production at $\mu
= 4~m_W$ and $\tan\beta = 2$ shown along with the SM background, with the
same convention as in Fig. 1.  The circles denote the fake LSD background.

\item[{\rm Fig.~6.}] The missing-$p_T$ $({p\!\!\!/}_T)$ distribution of
the LSD signals for
the $R$ conserving model are shown for gluino masses of 300, 600 and 1000
GeV with $p^\ell_T > 20$ GeV (solid lines) along with the fake
LSD background (dashed line).  The ${p\!\!\!/}_T$ distribution of the SM
background from $t\bar t$ is marginally softer than the latter, while that
from $b\bar b$ is extremely soft.

\end{enumerate}
\end{document}